\documentstyle[11pt,epsf]{article}    
\begin{document}    
\thispagestyle{empty}    
\rightline{UOSTP-01105}   
\rightline{{\tt hep-th/0108207}}    
    
\

\def\tr{{\rm tr}\,} \newcommand{\beq}{\begin{equation}}    
\newcommand{\eeq}{\end{equation}} \newcommand{\beqn}{\begin{eqnarray}}    
\newcommand{\eeqn}{\end{eqnarray}} \newcommand{\bde}{{\bf e}}    
\newcommand{\balpha}{{\mbox{\boldmath $\alpha$}}}    
\newcommand{\bsalpha}{{\mbox{\boldmath $\scriptstyle\alpha$}}}    
\newcommand{\betabf}{{\mbox{\boldmath $\beta$}}}    
\newcommand{\bgamma}{{\mbox{\boldmath $\gamma$}}}    
\newcommand{\bbeta}{{\mbox{\boldmath $\scriptstyle\beta$}}}    
\newcommand{\lambdabf}{{\mbox{\boldmath $\lambda$}}}    
\newcommand{\bphi}{{\mbox{\boldmath $\phi$}}}    
\newcommand{\bslambda}{{\mbox{\boldmath $\scriptstyle\lambda$}}}    
\newcommand{\ggg}{{\boldmath \gamma}} \newcommand{\ddd}{{\boldmath    
\delta}} \newcommand{\mmm}{{\boldmath \mu}}    
\newcommand{\nnn}{{\boldmath \nu}}    
\newcommand{\diag}{{\rm diag}}    
\newcommand{\bra}{\langle}    
\newcommand{\ket}{\rangle}    
\newcommand{\sn}{{\rm sn}}    
\newcommand{\cn}{{\rm cn}}    
\newcommand{\dn}{{\rm dn}}    
\newcommand{\tA}{{\tilde{A}}}    
\newcommand{\tphi}{{\tilde\phi}}    
\newcommand{\bpartial}{{\bar\partial}}    
\newcommand{\br}{{{\bf r}}}    
\newcommand{\bx}{{{\bf x}}}    
\newcommand{\bk}{{{\bf k}}}    
\newcommand{\bq}{{{\bf q}}}    
\newcommand{\bQ}{{{\bf Q}}}    
\newcommand{\bp}{{{\bf p}}}    
\newcommand{\bP}{{{\bf P}}}    
\newcommand{\thet}{{{\theta}}}    
\newcommand{\tauu}{{{\tau}}}    
\renewcommand{\thefootnote}{\fnsymbol{footnote}}    
\    
    
\vskip 0cm    
\centerline{   
\Large\bf Junctions of Supersymmetric Tubes    
}    
  
\vskip .2cm    
  
\vskip 1.2cm  
\centerline{   
Dongsu Bak$^a$ 
  and Sang-Woo Kim$^{b}$
}  
\vskip 10mm   
\centerline{ \it $^a$ Physics Department,   
University of Seoul, Seoul 130-743, Korea}   
\vskip 3mm   
\centerline{ \it $^b$ School of Physics, Seoul National University, 
Seoul 151-747, Korea}   
\vskip 0.3cm  
\centerline{\tt 
($\,$dsbak@mach.uos.ac.kr, eyed@phya.snu.ac.kr$\,$)}

\vskip 1.2cm         
\begin{quote}    
{
We begin by reviewing the noncommutative supersymmetric 
  tubular configurations in the matrix 
theory.  We  identify the worldvolume gauge fields,
the  charges  and 
the  moment of R-R charges carried by  
the tube. We also study the fluctuations around 
many tubes and tube-D0 systems.  
Based on the supersymmetric  tubes, we have constructed 
more general configurations that approach supersymmetric tubes
asymptotically. 
These include a bend with angle and a junction that connects
 two tubes to one. The junction may be interpreted as
a finite-energy domain wall  that 
interpolates U(1) and U(2) worldvolume
gauge theories.
We also construct a tube along which the noncommutativity scale changes.
Relying upon these basic units of operations, one may build 
 physical configurations corresponding to any shape of Riemann 
surfaces of arbitrary topology. 
Variations of the noncommutativity scale are allowed over 
the Riemann surfaces.
Particularly simple  
such configurations are Y-shaped junctions. 
} 
\end{quote}    
    

\newpage    
\section{Introduction}    
 
When external R-R field strengths are turned on, the lower dimensional 
D-branes may respond to the external fields like the dielectric material 
placed in an external electric field\cite{Myers}. 
A spherical D2 branes may be formed 
by D0's placed in an external R-R four form field strength and the  
worldvolume gauge theory becomes noncommutative.
Recently, it is found that the tubular D2-brane 
formed by D0's and strings may also be  
self supported from the collapse by its own  
worldvolume gauge fields\cite{Mateos,Klee}. 
This was originally realized in the 
Born-Infeld theory description of D2-branes\cite{Mateos}. 
The corresponding  
description from the matrix model was found thereafter including 
many supersymmetric tubes\cite{Klee}. 
This implies that there are no static 
forces between parallel  tubes of various sizes.  
The worldvolume gauge field theory is more accessible in this  
setting and the tube-D0 systems are described in terms of 
the solitonic excitations of the worldvolume gauge theory. 
The super D-helix is shown to be related to 
the supersymmetric tube  by 
the T-duality transformation\cite{Cho}. 
The supergravity  
solutions describing the tubular branes are identified recently  
and the dipole moment for the R-R four form field strength has been 
computed from the solutions\cite{Emparan2}.    
 
In this note, we shall first review the previous construction of tubes 
 from the matrix model. 
We shall identify the background magnetic field and electric field
on the tube.
The strength of electric field turns out  
to be critical while the strength of the background magnetic
component is arbitrary. We identify also the number of fundamental
strings stretched along the tube. The tube does not carry
net D2-brane charges. 
The moments carried by the 
tubes 
for the R-R four form field strength will be shown to agree to that  
of the supergravity solutions. We then study some fluctuations 
around the tube-D0 system and tubes with different radii.
 
We  move on the main subject of this note. Namely, we shall  
construct more general physical configurations of tubes. For certain 
set of initial   configuration to be physical, they ought to 
satisfy the Gauss law constraint. We will construct such  
configurations that approach  supersymmetric  tubes   
in the asymptotic region. The regions in which the BPS equations 
are violated will be kept finite with the excitation energy bounded. 
Within these conditions, we 
 ask most general configurations allowed by the system. 
 
We shall first show that 
the basic construct for building up  general tubes  
involves a junction of two tubes to one as well as a  
bending of tube. 
We shall also construct a tube along which the 
noncommutativity scale varies.
In a certain sense, these  basic  
operations are local excitations out of many asymptotically 
supersymmetric  tubes. 
In particular, the junction is a sort of local finite size  
domain wall interpolating U(2) noncommutative  
worldvolume gauge theory 
to U(1)  theory; a similar type of domain walls between
 different gauge groups was considered in Ref.~\cite{Bachas}.
The domain wall carries a 
finite energy. We shall not address the dynamics of such 
domain 
walls. 
 
To construct  general tubular configurations, 
we  combine the basic operations along the axis 
of the tube. 
This way one may construct, for example, a junction which connects 
a tube to $p$ tubes or  one tube splitting to two tubes and 
recombining 
to one along the $z$ direction. If one repeats such operations,  
one can build generic  Riemann surfaces of an  
arbitrary topology with varying noncommutativity scale.  
 
The plan of this paper is as follows. In section 2, we shall 
review the previous construction and identify the background 
electromagnetic field, the charges and
the moment 
of the tube. In section 3,  
the fluctuations around multiple 
tubes  will be discussed.  
A particular interest will be  
the pattern of the symmetry breaking of tube 
configurations with different radii. We shall further analyze 
the fluctuation  spectra of tube-D0 systems. This will correspond  
to the spectrum of 0-2 strings connecting D0's to the tube. 
In section 4, we shall study the basic construct for  
the generic configurations. These will include the joint 
of two  tube to one  as well as  the bending. 
We shall show that the noncommutativity scale may 
vary along the tube.
Utilizing these basic operations, one may construct
 arbitrary Riemann surfaces with varying noncommutativity scale.
Last section comprises concluding remarks.

\section{Supersymmetric Tube Solutions}    
 
To discuss the tube   
configurations, we begin with the matrix   
model  Lagrangian\cite{BFSS,Seiberg1} 
\begin{equation}    
L={1\over 2 R} \tr \left(\sum_I (D_0 X_I)^2  
+{1\over (2\pi \alpha')^2}
\sum_{I<J} [X_I,X_J]^2+ {\rm fermionic\ part}  
\right)   
\label{lag}    
\end{equation}    
where $I,J=1,2,\cdots 9$, $R=g_s l_s$ is the   
radius of tenth   
spatial direction and $\alpha'\equiv l_s^2$ is related to
 the eleven dimensional  
Planck length\footnote{There 
is a change from \cite{Klee} 
in the definition of the eleven dimensional Planck length  
$l_{11}$. The difference is a numerical factor by $(2\pi)^{1\over 3} 
l_{11}=l^{\rm old}_{11}$.} by $l_{11}=
(R \alpha')^{1\over 3}$. 
The scales $R$ and $2\pi\alpha'$ 
will be omitted below by setting them unity  
and we shall recover them whenever necessary.  
As is well known, this Lagrangian can be thought of   
describing $N$ D-particles if one takes all the dynamical  
variables as $N\times N$ matrices.   
  
Let us first describe relevant BPS equations   
we like to solve. For this, we shall turn on only first three   
components of the matrices $X_I$. Then the Gauss law reads  
\begin{equation}    
[X,D_0 X]+[Y,D_0 Y]+[Z,D_0 Z]=0\,.  
\label{gausslaw}    
\end{equation}    
Using the Gauss law  constraint, the bosonic part of the 
Hamiltonian can be rewritten as  
\begin{equation}    
H={1\over 2 } \tr \left( (D_0 X\pm i[Z,X])^2+(D_0 Y\pm i[Z,Y])^2+(D_0 Z)^2  
- [X,Y]^2+ 2 C_J  
\right)\  \ge\   \tr C_J  
\label{bound}    
\end{equation}    
where  $\tr C_J$  is the central charge  defined by   
\begin{equation}    
\tr C_J= \pm{i\over 2}\tr \sum^3_{i=1}\, [X_i\,, Z (D_0 X_i)+(D_0 X_i) Z]\,.   
\label{centralcharge}    
\end{equation}     
The saturation of the BPS bound occurs if the BPS equations  
\begin{eqnarray}    
&&  [X,Y]=0,\ \ \ \ \ \ \ D_0 Z=0\,,\nonumber\\  
&&\ D_0 X\pm i[Z,X]=0, \ \ \ D_0 Y\pm i[Z,Y]=0   
\label{bpsequations}    
\end{eqnarray}    
hold together with the Gauss law constraint. 
On the choice of gauge $A_0= 
{1\over \, 2\pi \alpha'}\, Z$, the  
BPS equations of the upper sign  imply that all the fields are  
static. Hence the system of equations  
 reduce to  
\begin{eqnarray}    
[X,Y]=0,\ \ \ \  [X,[X,Z]]+[Y,[Y,Z]]=0\,,  
\label{BPS}    
\end{eqnarray}   
where the latter comes from the Gauss law constraint. 
Before providing the   
representations of the algebra, let us count the remaining supersymmetries   
of the states specified by the nontrivial representation of  the   
algebra.   
The supersymmetric variation of the  fermionic  
coordinates $\psi$ is    
\begin{eqnarray}    
  \delta \psi =  \left(D_0 X^I\, \gamma_{I} + {i\over 2}[X^I,X^J]
\,\gamma_{IJ}\right)\epsilon +\tilde{\epsilon}\,,  
\label{susy}    
\end{eqnarray}    
where $\epsilon$ and $\tilde{\epsilon}$ are respectively 
real spinors of
16 components parameterizing total 32 supersymmetries.
Using the BPS equations, the invariance condition becomes
\begin{eqnarray}    
2(D_0 X\, \gamma_{1}+D_0 Y \,\gamma_{2})\Omega_+\epsilon 
+\tilde{\epsilon}=0\,,  
\label{susy1}    
\end{eqnarray}    
where the projection operators $\Omega_\pm$ are $(1\pm\gamma_3)/2$.
This is solved by $\epsilon=\Omega_-\eta$ and $\tilde{\epsilon}=0$
with $\eta$ arbitrary. 
The kinematical supersymmetries parametrized by $\tilde{\epsilon}$ 
are completely broken while half of remaining sixteen  
supersymmetries are left unbroken.  
Thus the configuration preserves a quarter   
of 32 supersymmetries of the matrix model, which is  
in agreement with the Born-Infeld  or the  
supergravity description of the tubes\cite{Mateos,Emparan2}.

Among the solutions of the BPS equations,  a tube 
or multiple tubes are described by the algebra, 
\begin{eqnarray}    
  [z,x]=il y, \ \ [y,z]=i l x\,, \ \   [x,y]=0,   
\label{bpssolutions}    
\end{eqnarray}    
with $X_i=x_i$.  
The length   
scale $l$ is the noncommutativity parameter of  
the worldvolume gauge theory.   
  
The algebra in (\ref{bpssolutions}) is realized as follows.  
Let us introduce variables $x_\pm$ by  
\begin{eqnarray}    
x_\pm= x\pm i y\,.  
\label{xpm}    
\end{eqnarray}    
%
It is clear that 
$x_-x_+ =x^2+y^2\equiv\rho^2$ is a   
Casimir operator.  
 
We are interested in the following irreducible  
representation  
of the algebra\cite{Chaichian},  
\begin{eqnarray}    
x_+|n\ket=\rho |n+1\ket\,,\ \ \   
z|n\ket=l (n+\epsilon) |n\ket \ \ \ \epsilon \in [0,1) \, , 
\label{represenation}    
\end{eqnarray} 
where $|n\ket\, (n\in {\bf Z})$ is the basis for the  
original matrix variables. 
Because the system is only invariant under the finite 
 translations to the  
$\hat{z}$ direction by $ml\, (m\in {\bf Z})$, there is this nontrivial 
parameter $l\epsilon$ characterizing the continuous translation 
modulo the finite translations.  
Below we shall set $\epsilon$ to zero 
 for simplicity.    

As $\rho^2$ is Casimir operator and can be regarded as a number, we 
can represent the $x_\pm$ with the angular variable as follows 
\begin{equation} 
x_\pm = \rho e^{\pm i \theta} 
\end{equation} 
with periodic Hermitian operator $\theta$. Then $e^{\pm i\theta}|n\ket 
= |n\pm 1\rangle$ and $[z,e^{\pm i \theta}] = \pm l e^{\pm i\theta} $. 
It is obvious that our BPS configuration describes a noncommutative 
tube of radius $\rho$ in three dimensions.  The coordinates 
$(\theta,z)$ on the tube would be noncommutative. 
 
Any well-defined operator   can be presented as  
\begin{eqnarray}    
f(z,\theta) =\sum^\infty_{n=-\infty}\int^{\pi\over l }_{-{\pi\over l}}  
{dk\over 2\pi} \tilde{f}_n (k)  
e^{i n \theta+ik z}\,.  
\label{function}    
\end{eqnarray}   
The range of $k$ is determined by the fact that the $z$ operator has 
discrete eigenvalues. Also any operator can be represented as  
$ f= \sum_{n,m} f_{nm}|n\ket\bra m|$ in the matrix 
theory. Two representations are related by 
\begin{eqnarray}    
f_{nm}=\int^{\pi\over l}_{-{\pi\over l}}  
{dk\over 2\pi} \tilde{f}_{n-m} (k)  
e^{ {i l\over 2} (n+m)k}   
\,.  
\label{component}    
\end{eqnarray}   
%

The multiplication of operators on the  noncommutative tube is well 
defined. One may instead introduce 
the $*$-product of ordinary functions 
on the corresponding commutative  tube.  In the 
Fourier representation of ordinary functions, their $*$-product should 
lead to the  Fourier  
representation which we would get as the product of operators. Thus, 
the $*$ product of two ordinary functions $g$ and $h$ would be 
\begin{eqnarray}    
g* h= \left[e^{{il\over 2}(\partial_\theta \partial_{z'}-  
\partial_z \partial_{\theta'}  
)}  g(\theta, z) h(\theta',z')\right]_{\theta=\theta', z=z'}\,.   
\end{eqnarray}   
In addition, the spatial integration $\int d\theta dz$ on the tube 
corresponds to $2\pi l \tr $.  Since above $*$-product implies that 
$\theta * z- z* \theta =i\,l$, the minimal area is in a rough sense 
given by $2\pi  \rho\, l$.  Since the circumference of 
the tube is $2\pi \rho$, one may regard the noncommutativity scale $l$ 
as a  minimal distance in the $z$ direction. Indeed the 
discreteness of the spectrum of $z$ is consistent with this 
observation.  Moreover, $1/(2\pi \rho\, l)$ corresponds 
to the area density of 
the the constituent D0-branes. 
The total number of D0-branes is $ N=\tr I= {1\over 
2\pi \rho \,l}\int dz  d\theta \rho = L_z/ l$, where $L_z$ is the length 
of the tube in the $z$-direction.  
Hence the D0 brane density per unit length in the z-direction   
is $1/l$.  
 
At this point, let us work out the relation of the supersymmetric tube  
 to the string theory 
configuration. For this, we shall work out the  
charges involved, the worldvolume electromagnetic field 
and the dipole moment of the tube for the R-R four 
form field strength.  
For the comparison to string theory, we shall restore  
$R$ and $2\pi \alpha'$ in this part.

The energy for the  
supersymmetric tube is given by the central charge,  
\begin{equation}    
E= \tr C_J= {l^2\over g_s l_s (2\pi l_s^2)^2} \tr x_- x_+  = {l^2 \rho^2  
\over (2\pi)^2 g_s  l_s^5}\tr I={1 
\over (2\pi)^2 g_sl_s^3}{l \rho\over 2\pi l_s^2}  \int dz  d\theta\,\rho\,.   
\end{equation}     
We compare this with  
the total energy  of D2-brane in the M-theory point of view. 
Namely, we expand the membrane Hamiltonian 
$\sqrt{p^2_{11}+E^2_{\rm M2}}$ to the leading order by  
$p_{11}+ {E^2_{\rm M2}\over 2p_{11}}$ where $p_{11}=N/R$ is the 
momentum along the tenth spatial direction and $E^2_{\rm M2}$ is the 
total energy   squared carried by the membrane except  $p^2_{11}$. 
Thus the energy of the matrix model will be compared to 
\begin{equation}    
E={E^2_{\rm M2}\over 2p_{11} }\,.   
\label{energyD}    
\end{equation}    
Hence, the membrane  energy $E_{\rm M2}$ is evaluated as 
\begin{equation}    
E_{\rm M2}=  
 {1\over  (2\pi)^2 g_s\, l_s^3} (2\pi \theta_{\rm nc} \tr I)\sqrt{2} 
=  {1\over  (2\pi)^2 g_s\, l_s^3} V_T \sqrt{2} 
 \,,   
\end{equation}     
where $V_T$ is the spatial volume of the tube and 
$\theta_{nc}=l\rho$.  
The  
factor $\sqrt{2}$ is due to
the kinetic contribution that  
balances the one from the 
quartic potential.

For the worldvolume electromagnetic fields, let us first  
note that 
\begin{equation}    
[\theta, z]\sim i\,l\,.   
\end{equation} 
As in the case of the planar noncommutative  D2-brane, 
we define the worldvolume gauge fields  
by\footnote{Due to the compactness in the $\theta$-direction, 
only $x+i y=\rho e^{i\theta}$  
or $X+iY\equiv \rho e^{i\Theta}$ with $X^2+Y^2=\rho^2$ 
are  
well defined. However, for the clarity of argument,  we  
shall ignore this issue.} 
\begin{equation}    
\Theta=\theta + l A_z\,, \ \ \  Z=z- l A_\theta\,.   
\end{equation} 
Using $[\theta, \cdot]=il\partial_z$ and  
$[z, \cdot]=-il\partial_\theta$, we are led to  
\begin{equation}    
[\Theta, Z]= il^2 \left({1\over l}+ F_{\theta z}\right)\,.   
\label{magnetic}    
\end{equation} 
Thus it is clear that the worldvolume background magnetic  
field is 
\begin{equation}    
B={1\over l}\,.   
\label{bakmagnetic}    
\end{equation}   
Further noting $D_0\Theta= -l E_z$, the  
electric field on the tube may be evaluated as 
\begin{equation}    
E_z= -{1\over l} D_0 \Theta \sim {1\over l}(  
 \partial_X\Theta\,\, D_0 X +  \partial_Y\Theta\,\, D_0 Y)  
=-{1\over l \rho^2} ( X D_0 Y -  Y D_0 X) 
\,,   
\label{bakelectric}    
\end{equation} 
where we have ignored the operator ordering problem. 
Using the explicit tube solution, one finds that 
\begin{equation}    
E_z={1\over 2\pi \alpha'} 
\,,      
\end{equation} 
which is in agreement with that of the Born-Infeld description of 
the  supersymmetric tube in \cite{Mateos}.
 
The momentum conjugated to the $\theta$ coordinate 
 carried by the tube will be  
\begin{equation}    
\Pi_\theta= {1\over g_s l_s} \rho^2 D_0\Theta={1\over g_sl_s}
( XD_0 Y-Y D_0 X) 
= -{\rho^2 l\over 2\pi g_s l_s^3}I 
\,.      
\end{equation}  
Using the semiclassical Bohr-Sommerfeld quantization rule, 
\begin{equation}    
\oint \Pi_\theta d\theta= 2\pi N_s \ \ \ (N_s \in {\bf Z}) 
\,,      
\end{equation}   
we conclude that  
\begin{equation}    
{l \,\,\rho^2 \over 2\pi g_s l_s^3}= N_s 
\,,      
\end{equation}  
which implies that the radius of the tube is 
quantized semiclassically. The momentum $\Pi_\theta$  
in fact corresponds to the 
displacement current density in the $z$-direction 
and the integer $N_s$ counts the number of fundamental 
strings stretched along the $z$ direction. 

The total angular momentum along the $z$-axis  
\begin{equation}    
J_{\rm tot}={1\over g_s l_s}\tr(X D_0 Y-Y D_0 X )= - 
{l \rho^2\over 2\pi g_s l_s^3} N =-N_s N\,,   
\end{equation}     
is an integer and saturates the bound  
\begin{equation}    
|J_{\rm tot}|\le N_s N\,.   
\label{abound}    
\end{equation}
(As will be shown below,  tube-D0 systems
are the examples where  
the bound is not saturated. See also the case of many tubes.)       
In fact this may be compared to the bound on the angular momentum 
per unit length in the z-direction in \cite{Mateos}, 
\begin{equation}    
|j|\equiv |J_{\rm tot}/L_z|\le |\Pi_\theta\, B|\,.   
\end{equation}

Let us now turn to the problem of computing the dipole moment 
which generates the R-R four form field strength. 
For this purpose, we will use the nonabelian Chern-Simons  
couplings of D-particles to the R-R gauge fields, 
\begin{equation}    
S_{\rm CS}= 
\mu_0 \int dt \tr \left( C^{(1)}_t 
+C^{(1)}_{I} D_t \phi^I+{i\lambda\over 2}  
C^{(3)}_{t\,IJ}[\phi^J,\phi^I]+{i\lambda^2\over 3} 
\phi^I\phi^J\phi^K F^{(4)}_{t\,IJK}+{\rm h.o.t.}\right) 
\,,   
\end{equation}     
where $\mu^{-1}_p=(2\pi)^p g_s\, l_s^{p+1}$, $\lambda=2\pi \alpha'$, 
$X^I=2\pi \alpha' \phi^I$ and $F^{(p+1)}$ is the field strength 
corresponding to the R-R p-form potential, $C^{(p)}$.  
(See  
Ref. \cite{Myers} for the details.)  
The second and the third terms do not contribute to  
the interaction since $X^I$ and $[X^I,X^J]$ are traceless on  
the tube solution. In particular, the vanishing of the third term 
 implies there is no  
$D2$-brane charge carried by the tube configurations. 
This is as expected since such  Dp-branes  
involving a compact direction of  trivial cycle cannot carry  
net Dp-brane charges. 
 
The first term is nonvanishing and tells us that 
the number of D-particles is given by $\tr I$. the  
forth term is  
related to the dipole moment for the four form field strength. 
Noting 
\begin{equation}    
S^{\rm dipole}_{\rm CS}= 
{i\mu_0\lambda^2\over 3} \int\, dt\, \tr 
\phi^I\phi^J\phi^K F^{(4)}_{t\,IJK}= 
-{1\over 3}{\rho\over(2\pi)^2  g_s \,\,l_s^3}  
\int dt dz d\theta \rho  
F^{(4)}_{txyz} 
\,,   
\end{equation}       
we conclude that the dipole moment density of the tube is 
\begin{equation}    
d_2= {1\over 3} 
{\rho\over (2\pi)^2 g_s\,\, l_s^3}={1\over 3}\,\mu_2\, \rho\,.  
\end{equation}     
 
\section{Fluctuation Spectra of Multiple Tubes  
and Tube-D0 Systems} 
 
In Ref. \cite{Klee}, the  
solutions for the multiple tubes and the  
tube-D0 are also constructed. We shall analyze the   
fluctuation spectra around these solutions. Due to the  
complexity of geometries, the fluctuation spectra are also 
quite involved. Hence we shall restrict our attention to  
 two kinds of simple configurations. First  
is the symmetry breaking pattern of 
the concentric two tubes with different radii. The other is  
the spectra of tube-D0 strings when D0's are located  
at  the axis of the tube.  
 
First let us recall that the BPS solution  
describing  many parallel   tubes are given by 
\begin{eqnarray}   
X_+ &=& \sum_{a=0}^{p-1} \rho_a \sum_{n=-\infty}^{\infty} 
 |(n+1)p+ a\rangle\langle np+a| 
+  \sum^{p-1}_{a=0} \xi_a 
 \sum^\infty_{n=-\infty}|np +a\ket\bra np+a|\, , \nonumber  \\ 
Z\ \,  &=& \sum_{a=0}^{p-1} l_a \sum_{n=-\infty}^\infty n\,|n p+a\ket \bra 
np+a|  
\, ,    
\end{eqnarray}   
where $X_+=X+iY$  and 
$p$ is the number of tubes.
Here $\rho_a$ is for the radius of each tube, $l_a$ is for the 
noncommutative parameter of each tube, and $\xi_a$ is for the position 
of the center of each tube in $(x,y)$ space.  
 Of course one may add the position 
along the other dimensions.  
When all the noncommutativity parameters agree i.e. $l_a=l$, this 
background makes the worldvolume theory being a $U(p)$ noncommutative 
gauge theory. The $U(p)$ basis can be constructed by writing 
$|np+a\ket \bra mp +b|=|n\ket'\bra m|' T_{ab}$.   Here $|n\ket'$ is 
interpreted as a new basis for the space while $T_{ab}$ generates 
$U(p)$ algebra\cite{Park}. The total angular momentum for the many 
tubes is evaluated as $J^{\rm tot}=-\sum_a l_a\rho_a^2 \,\,\tr' I'$
where $I'\equiv \sum_n |n\ket'\bra n|'$ and
$\tr'$ is over the basis spanned by $|m\ket'\bra n|'$. 
Since
$\sum_a l_a\rho_a^2 $ is the total number of strings through the tubes
and $\tr' I'= N^{\rm tot}/p$, the angular momentum is related to
the numbers of strings and D0's by\cite{Emparan2}
\begin{eqnarray}   
|J^{\rm tot}|={1\over p}N_s^{\rm tot}N^{\rm tot}\,.
\end{eqnarray} 
 
To study the symmetric breaking involved with the multiple 
tubes, we shall consider the two concentric tubes with the same  
noncommutativity parameter but different radii. 
The two tubes are described by
\begin{eqnarray}   
x_+&=& \sum_{a=0}^{1} \rho_a \sum_{n=-\infty}^{\infty} 
 |2(n+1)+ a\rangle\langle 2n+a|\, , \nonumber  \\ 
z\   &=& l \sum_{a=0}^{1}  \sum_{n=-\infty}^\infty n\,|2n +a\ket \bra 
2n+a|  
\, .    
\end{eqnarray}   

Before discussing the fluctuation spectra of the above  
tubes, let us consider the equations satisfied by the transverse scalars 
 in the background of a tube.  
Turning on only one component, $X_4$, the equation of motion  
becomes 
\begin{equation}    
[\partial_t-iz,[\partial_t-iz,X_4]]+  
[x_i,[x_i,X_4]]=0 \,.   
\label{scalarY}   
\end{equation}   
First, let us rewrite the   
equation in the continuum limit. To this end,  
we note  
\begin{eqnarray}    
&&[z,\,\,\cdot\,\,]= -i l \partial_\theta\,,  
 \nonumber\\    
&& [x,\,\,\cdot\,\,]= -ily \partial_z +O(l^2)\,,\ \ \ 
[y,\,\,\cdot\,\,]= il x \partial_z +O(l^2)\,,  
\label{approximationY}    
\end{eqnarray}    
where the first line is exact and, in the second line, 
we have ignored order $l^2$ terms appearing due to the 
discreteness of $z$-direction. 
Using these relations, the above  scalar equation can be written as  
\begin{equation}    
\left(\partial_t-{l\over 2\pi \alpha'}\partial_\theta\right)^2X_4-  
\left({l\,\rho \over 2\pi \alpha'}\right)^2 
\left({1\over  \rho^2} \partial_\theta^2 +\partial_z^2\right)  
X_4  
+O(l^3)  
=0 \,,  
\label{waveY}    
\end{equation} 
where we have restored our units.  
Introducing  
$\theta'=\theta 
+{l\over 2\pi \alpha'}\,t$,  
the above equation becomes  
a standard wave equation  on a cylinder, 
\begin{equation}    
\partial_t^2X_4 - 
\left({l\,\rho \over 2\pi \alpha'}\right)^2 
\left({1\over  \rho^2} \partial_{\theta'}^2 +\partial_z^2\right)  
X_4 
+O(l^3)  
=0 \,.  
\end{equation} 
The extra terms in the wave equation (\ref{waveY}) are produced 
by the presence of the background electric field.  
As shown in \cite{Klee}, the continuum limit of the  
worldvolume gauge theory on the tube produced by 
the fluctuation of $X,Y,Z$ 
is the noncommutative U(1) Yang-Mills-Higgs system  
with a peculiar Chern-Simons term. Here the  
coordinate $\theta'$ were used and the geometry 
is flat and noncommutative. The origin of the Chern-Simons 
term in the worldvolume theory is not clear. 
 
Let us now turn to the case of the two concentric tubes with  
different radii.  
Here we expect that an analog of the spontaneous  
symmetry breaking occurs. Namely, the breaking of the $U(2)$ 
symmetry to $U(1)\times U(1)$ from the view point of the 
worldvolume gauge theory. However, the symmetry breaking pattern 
is more involved. 
 For example, 
the wave equations for each $U(1)$ component of 
 the transverse  
scalar are dependent upon the scale $\rho_{a}$ explicitly 
as we can see in (\ref{waveY}). The effective metric and the 
Yang-Mills couplings are also dependent upon the radii. 
This behavior 
is quite different from that of the  Yang-Mills theory 
on the parallel flat Dp-branes  where the effective Yang-Mills 
coupling 
or the metric do not depend on the transverse separations. 
 
The W-boson corresponds to the strings connecting one tube 
to the other. The mass produced by the radial separation 
may be found by considering the quadratic fluctuations around 
background two-tube configuration. Indeed, one may explicitly 
verify that the extra mass squared term produced by the separation  
is proportional to the radial separation squared, 
\begin{equation}    
(\Delta \rho)^2\equiv |\rho_{0}-\rho_1|^2 \,.   
\end{equation}    
 
Let us now discuss the probe of the supersymmetric tube  
using D0 solitons.  
In Ref.\cite{Klee},  multi soliton solutions describing 
D0's are constructed.   
For this we  introduce a shift operator defined by 
\begin{eqnarray}    
S = \sum^\infty_{n=0} |n+m\ket \bra n|  
+ \sum_{n=-\infty}^{-1} |n\ket \bra n|\,.  
\end{eqnarray}      
It satisfies the relations  
\begin{eqnarray}    
S S^\dagger= I-P\,,\ \ \   
S^\dagger S =I\,,  
\end{eqnarray}       
where the projection operator $P$ is defined by $ 
P=\sum^{m-1}_{a=0} |a\ket \bra a| $.  
The general soliton solutions including the moduli parameters 
are given by 
\begin{equation}  
 \bar{X}_i=S \, x_i S^\dagger + 
\sum^{m-1}_{a=0}\lambda_i^{a}|a\ket\bra a| \, , \,\,\,\,  
 \bar{X}_s= \sum^{m-1}_{a=0}\varphi_s^{a}|a\ket\bra a|\,,  
\label{tubed0}    
\end{equation} 
with the index $s$ referring to the transverse scalar $X_4$ 
to $X_9$.  
Unlike the solitons of the noncommutative Yang-Mills theory 
describing a 
planar D2-brane\cite{Gopakumar,Bak,Aganagic,Harvey,Witten},   
the solutions we constructed here are  
BPS saturated states of eight supersymmetries. 
Namely, the solution satisfies the original BPS equations 
and no further supersymmetries are broken by the presence of the
 solitons. 
The moduli are describing the positions of D0 branes 
in the 9-dimensional space. The appearance of moduli further 
support the view point that the configurations 
are describing not holes in tubes but extra  D0-branes 
that may even fly off the tube. 
Since there are extra D0's, the bound in (\ref{abound}) is no 
longer saturated. Namely the total angular momentum is smaller
than the number of strings multiplied by the total number of
D0's. 
 
For simplicity of the analysis in studying 
 the fluctuation around this background, we shall set all the moduli 
parameters to zero. This corresponds to D0's located at 
the origin of the nine spatial  dimensions, which is at the axis 
of the supersymmetric tube. (For the gauge invariant 
 positions of noncommutative solitons, see Ref. \cite{Bak}.) 
 
To study dynamics of the system, we choose the gauge $A_0=Z$ and  
consider  small fluctuations of 
the matrix variable around the solution (\ref{tubed0})  
\begin{eqnarray}    
&& Z=\bar{Z}+P\delta Z P+P\delta Z \bar{P}+\bar{P}\delta Z P+ 
\bar{P}\delta Z \bar{P}\,,\nonumber\\  
&& X_+=\bar{X}_+ +P\delta X_+ P+P\delta  X_+\bar{P}+\bar{P}\delta  X_+ P+ 
\bar{P}\delta  X_+ \bar{P}\,,\nonumber\\ 
&& X_s=
P\delta X_s P+P\delta  X_s\bar{P}+\bar{P}\delta  X_s P+ 
\bar{P}\delta  X_s \bar{P} 
\end{eqnarray}     
where $\bar{P}=I-P$. 
We shall insert this expression to the original matrix model  
Lagrangian and find a new effective  Lagrangian 
to the quadratic order of the fluctuations. As in the case  
of the Abelian-Higgs system\cite{Bak} or  
D2-D0 system\cite{Aganagic}, the linear fluctuation of  
$\bar{P}\delta X_I \bar{P}$  are described  by the original  
worldvolume dynamics  of the tube without $D0$'s. This would 
corresponds to the dynamics of tube-tube strings. On the other  
hand, the linear fluctuation of $P\delta X_I P$ are described by the 
original matrix model with $m\times m$ matrix variables. This implies 
that the fluctuations describe the 0-0 strings connecting  
$m$ D0's. These two kinds of fluctuations decouple from the rest 
at the order of linear fluctuation. 
The remaining fluctuations would correspond  the tube-0 strings. 
In order to find the spectra of these strings, we shall find the 
corresponding equations of motion to the linear order. For this  
purpose, we will use the following matrix variables, 
\begin{eqnarray}    
&& T_1 ={1\over \rho}P\delta X_+\bar{P} \bar{X}_- S,\nonumber\\ 
&& T_2 ={1\over \rho}P\delta X_-\bar{P} \bar{X}_+ S,\nonumber\\ 
&& K= P \delta Z\bar{P}  S 
\end{eqnarray}     
where all of these are complex $m\times \infty$ matrices, e.g.  
$M_{an}\,(a=0,\cdots, m-1\ {\rm and}\ n\in {\bf Z})$. 
With straightforward computations, one finds the following set of 
linearized equations, 
\begin{eqnarray}    
&&{d\over dt} \left(2\dot{K}-i\rho(\dot{T_1}+\dot{T_2})\right)=0, 
\nonumber\\ 
&& \ddot{T}_1+{\rho^2\over 2}(T_1-T_2)-i\rho \dot{K} 
+2 i \dot{T}_1 (z-l )=0 
\nonumber\\ 
&& \ddot{T}_2+{\rho^2\over 2}(T_2-T_1)-i\rho \dot{K} 
+2 i \dot{T}_2 (z+l )=0\,. 
\end{eqnarray}     
There is also one equation from the Gauss law but it is  
not independent.
The  
equations above show that there is no mixing between  
matrix components with different indices. The first equation can  
be easily integrated and one obtains  
$2\dot{K}=i\rho({T_1}-{T_2})$. (The integration constant  
is set to zero.) Using this relation, we eliminate $K$ from  
above equations and get 
\begin{eqnarray}    
&& (\ddot{T}_1)_{an} 
+2 i l(n-1) (\dot{T}_1)_{an} +\rho^2 (T_1)_{an} =0\nonumber\\ 
&& (\ddot{T}_2)_{an} 
+2 i l(n+1) (\dot{T}_2)_{an} +\rho^2 (T_2)_{an} =0 
\end{eqnarray}     
We now introduce new matrix variables by 
\begin{eqnarray}    
&& U =P\delta X_+\bar{P} S,\nonumber\\ 
&& V =P\delta X_-\bar{P} S, 
\end{eqnarray}      
and the equations of motion for these now become 
\begin{eqnarray}    
&& \ddot{U}_{an} 
+2 i ln \dot{U}_{an} +\rho^2 U =0\nonumber\\ 
&& \ddot{V}_{an} 
+2 i ln \dot{V}_{an} +\rho^2 V 
=0\,. 
\end{eqnarray}       
For the transverse coordinates $X^s$, their  
fluctuations are governed by the same equations, i.e. 
\begin{eqnarray}    
(\partial_t^2 
+2 i ln \partial_t +\rho^2) T^s_{an} =0\,, 
\end{eqnarray}     
where we define $T^s\equiv P\delta X^s\bar{P} S $. 
The corresponding angular frequency for all these modes is 
\begin{eqnarray}    
\omega_n^\pm={1\over 2\pi \alpha'} 
\left(ln \pm\sqrt{\rho^2+(ln)^2}\right)\,, 
\end{eqnarray}    
where we have restored the units. 
The first term in the parenthesis is again  
coming from the fact that there is the nonvanishing  
electric field or angular momentum carried by the  
tube configuration. The spectra are independent  
of the D0-brane index $a$ as it should be. 
The second contribution, $M_n\equiv  
{1\over 2\pi \alpha'}\sqrt{\rho^2+(l\,n)^2}$, in the frequency  
is also the one expected. It corresponds to the mass of  
the string stretched between D0's at the origin to the  
point on the tube at $z=ln$. Namely, the mass for such string 
is given by the string  tension multiplied by the stretched length, 
which agrees precisely with $M_n$. 
Thus we conclude that the geometry seen by D0-probe is a tube 
extended in the z-direction 
with radius $\rho$. 
Finally, we like to compute the contributions of above modes 
to the Hamiltonian of the system. But evaluating the contribution to 
 $H$ requires the terms in the second order in the fluctuation 
due to the couple to the background energy. Instead, we will 
evaluate $\tilde{E}\equiv H-\tr  C_J$. Since $C_J$ is conserved, the  
combination 
is also conserved. The straightforward evaluation is given by 
\begin{eqnarray}    
\tilde{E}={1\over 2} 
\sum_{an} \left(\dot{U}_{an}^2 +\dot{V}_{an}^2 + 
\rho^2 
(U^2_{an}+V^2_{an})+  
\sum_s\Bigl[(\dot{T}^s_{an})^2 + 
M_n^2 (T^s_{an})^2\Bigr]\right) 
\,. 
\end{eqnarray}   
Here $K$ is eliminated again and the above form explicitly shows 
that there are indeed only two independent oscillators; one may  
regard $K$ as a gauge degree 
of freedom. The disparity between the transverse modes and $U,V$ 
is mainly due to subtraction of the  
central charge from the Hamiltonian.

\section{Junctions and Bends of Tubes} 
 
So far we have described the characteristics of  
the supersymmetric tubes by looking at the charges 
or worldvolume  fluctuations.  
 In this section  
we like to describe the junctions of three tubes  
that approach 
asymptotically the supersymmetric  tubes extended along  
the same direction. 
Then such junctions may be considered as local  
excitations in a  
certain sense because the remaining supersymmetries are  
broken locally at the junctions. If one splits, for example, a 
tube to two concentric tubes of equal radii, the junction will  
interpolate  
     the U(1) noncommutative worldvolume 
theory on one tube  to the U(2) noncommutative gauge theory 
on the concentric tubes. The interpolation is local because it 
occurs along the $z$ directions and the area of the excited
region is 
finite. The excitation of energy of this domain  
wall is also finite. 
In short, it is a domain wall  
interpolating between  U(1) and U(2) noncommutative  gauge  
theories.  
 
The second element of deformation will be a bend of tube 
that approaches asymptotically  supersymmetric  
configurations. In particular, we shall construct a bend of tube 
that makes an  asymptotic angle.  
The region where the  
bending occurs will be  local. We call this type of  
configuration as bend with angle $(\kappa,\nu)$ where 
the angles are defined as follows. If the lower part of the tube is  
extended in the z-direction, $\kappa$ is the angle of the upper part of the 
tube with  
z-axis and $\nu$  the azimuthal angle. 
By performing  double bending  operations,  one may construct  
a bend of tube with displacement, which connects a tube  
of the $z$ direction  centered at $(\xi_x,\xi_y)$  
in the $(x,y)$ space 
 to the other tube at $(\xi'_x,\xi'_y)$. 
This of course can be done with the bending of angle ($\kappa$,0) at  
some point of tube and the other bending of angle ($\kappa\,,\pi$) at  
another point.  
We also construct such configuration directly and  
study more details of parameters involved. 
The remaining basic construct is  a tubular configuration along which
 the 
noncommutativity scale changes.

 Combining  these 
configurations, one may construct  arbitrary 
Riemann surfaces. For example, a tube with one hole can be easily 
constructed; for two tubes that are appropriately bended, 
one joins 
the lower ends as well as the upper ends
by the junctions. 
 
Let us first consider the construction of the simple junction. 
Here we shall use the gauge $A_0=Z$.  
The tube for $z \rightarrow -\infty$  approaches the  
supersymmetric tube with radius $\rho$ whereas,  
for $z \rightarrow \infty$, it approaches two tubes with radii $\alpha$  
and $\beta$. 
The ansatz for the solution is then 
\begin{eqnarray}    
&& Z\,\,={\cal Z} +l \sum^{-(b+1)}_{n=-\infty} n|n\ket \bra n| 
+l \sum^{\infty}_{n= a}n \Bigl(|2n\!+\!1\ket \bra 2n\!+\!1|+ 
|2n\!+\!2\ket \bra 2n\!+\!2|\Bigr) 
\,,  
\nonumber\\   
&&  X_+={\cal X}_+ + \rho \sum^{-(b+1)}_{n=-\infty} |n\!+\!1\ket \bra n| 
+ \sum^{\infty}_{n= a} \Bigl(\alpha |2n\!+\!1\ket \bra 2n\!-\!1|  
+\beta |2n\!+\!2\ket \bra 2n|
\Bigr) \,,  
\label{ansatz}    
\end{eqnarray}    
where ${\cal Z}$ and ${\cal X}_+$ are  
$(2a+b+1)\times (2a+b+1)$ matrices. The basis of these finite matrices 
is spanned by $|\!-\!b\ket,|\!-\!b\!+\!1\ket, 
\cdots,|2a\!-\!1\ket, |2a\ket$.  
We insert this ansatz to the Gauss law and find an equation 
\begin{eqnarray}    
[{\cal X}_+,[{\cal X}_-, 
{\cal Z}]] + [{\cal X}_-,[  {\cal X}_+,{\cal Z}]]   
=2l\alpha^2 |2a\!-\!1\ket \bra 2a\!-\!1|+2l\beta^2 |2a\ket \bra 2a| 
-2l\rho^2 |\!-\!b\ket \bra \!-\!b|
\,.  
\label{master}    
\end{eqnarray}    
By the Gauss law, the boundary conditions  
for ${\cal Z}$ and ${\cal X}_+$ are given as  
\begin{eqnarray}    
\bra -b\, |{\cal Z}|\!-\!b\ket = -lb\,,\ \ \  
\bra 2a\!-\!1| {\cal Z}  |2a\!-\!1\ket = 
\bra 2a| {\cal Z}  |2a\ket= l a\,,  
\label{condition1}    
\end{eqnarray}    
and  
\begin{eqnarray}    
&&\Bigl({\cal Z}+l(b\!-\!1)\Bigr){\cal X}_+|\!-\!b\ket =0\,,\nonumber\\ 
&& \Bigl({\cal Z}-l(a\!-\!1)\Bigr){\cal X}_-|2a\!-\!1\ket  = 
\Bigl({\cal Z}-l(a\!-\!1)\Bigr){\cal X}_-|2a\ket=0\,.   
\label{condition2}    
\end{eqnarray}   
Taking trace of Eq.~(\ref{master}),  
one obtains the relation 
\begin{eqnarray}    
l \rho^2=l \alpha^2+ l\beta^2\,.   
\end{eqnarray}   
Since $l\,{(radius)}^2 
\sim {\rm \#\ of\  
fundamental\ strings}$, the above relation implies that the number of  
fundamental strings is preserved through the junction. 
Finding most general solutions of the above  is  
quite involved. 
Let us  work out the  
simplest case, i.e. $a=1$ and $b=0$. 
In this case, the $3\times 3$  
matrix ${\cal Z}$ is fixed completely as 
\begin{eqnarray}    
{\cal Z} = l|1\ket\bra 1|+ l |2\ket\bra 2| 
\label{z1}    
\end{eqnarray}    
by the condition (\ref{condition1}). Imposing the  
condition in (\ref{condition2}) on ${\cal X}_+$ and solving  
(\ref{master}), one finds that 
\begin{eqnarray}    
{\cal X}_+ = |0\ket\bra \psi|+  |\phi\ket\bra 0| 
\label{x1}    
\end{eqnarray}    
with 
\begin{eqnarray}    
|\psi\ket = \alpha \cos \varphi |1\ket +\beta\sin\varphi |2\ket 
\,,\ \ |\phi\ket = -\alpha \sin \varphi |1\ket +\beta\cos\varphi 
|2\ket 
\,. 
\end{eqnarray}    

\begin{figure}
\epsfxsize=4.6in
\centerline{
\epsffile{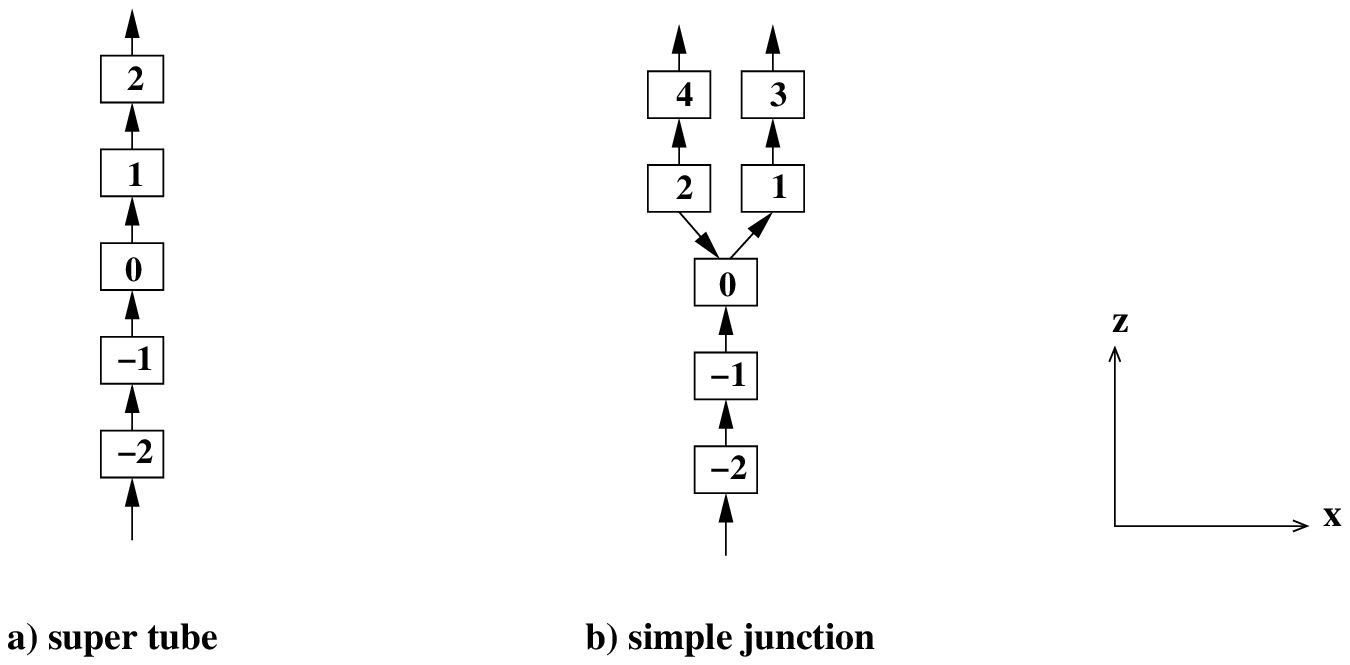}
}
\vspace{.1in}
{\small Figure~1:~A skeleton graph for the 
supersymmetric tube is depicted first. Here the box with number $i$ 
denotes the basis $|i\ket$ and 
its position corresponds to the eigenvalue
of the diagonal matrix $Z$. The arrow 
from the box $i$ to the box $j$ indicates  that 
there is a nonvanishing component of $|j\ket\bra i|$ in $X_+$. The second
figure is for the simple junction in (\ref{z1}) and (\ref{x1}) with 
$\varphi=0$.}
\end{figure}

This will be the generic configuration (within the ansatz) 
which is consistent with the Gauss law.  
(Here we do not include some trivial phase factors that  
amount to the gauge parameters.) 
It does not 
satisfy the equations of motion but serves as a valid  
initial  configuration. Staring from this   
configuration, 
the system will eventually   evolve  
toward lower energy configurations. We call such initial  
configuration as  physical one. 
Let us compute the energy of the junction
which is, of course, conserved during the  
later time evolution. 
For this purpose, we shall compute the following 
energy, 
\begin{eqnarray}    
\tilde{E}=H-\tr C_J 
\equiv {1\over 2} \tr \left(\partial_0 X^i \partial_0 X^i -[X,Y]^2\right)     
\,. 
\label{h-j}    
\end{eqnarray}    
For the above  
initially static configuration, the  energy 
is evaluated as 
\begin{eqnarray}    
\tilde{E}= 
(\alpha^2 \sin^2\varphi +\beta^2 \cos^2\varphi)^2 
+{1\over 4} (\alpha^2+\beta^2) 
(\alpha^2\sin^2\varphi+\beta^2\cos^2\varphi)  
+{1\over 4}(\alpha^4 \sin^2\varphi + 
\beta^4 \cos^2\varphi)   
\,. 
\end{eqnarray}    
When $\alpha \ge \beta$, the 
energy is minimized if $\sin\varphi=0$; the energy at the  
minimum becomes 
\begin{eqnarray}    
\tilde{E}_{\rm min}={1\over 16 \pi^2 g_s l_s^5}\left( 5   \beta^4 + 
\rho^2\beta^2\right)   
={g_s l_s\over 4\, l^2} N_\beta(5 N_\beta+N_\rho ) 
\,, 
\end{eqnarray}  
where $N_\beta$ and $N_\rho$ are respectively the numbers 
of fundamental strings through tubes of radii $\beta$ and  
$\rho$. Note that the expression is  
valid 
only when $\sqrt{2}\beta \le\rho $ due to 
the condition $\beta \le \alpha$.
The corresponding configuration is depicted in Figure~1.
We shall comment more on the energetics of  
the configuration. 
 
Let us now consider 
the bend without asymptotic angles. 
The corresponding ansatz will  be 
\begin{eqnarray}    
&& Z\,\,={\cal Z} +l \sum^{-(b+1)}_{n=-\infty} n|n\ket \bra n| 
+l \sum^{\infty}_{n= a+1}n |n\ket \bra n| 
\,,  
\nonumber\\   
&&  X_+={\cal X}_+ + \rho \sum^{-(b+1)}_{n=-\infty} |n\!+\!1\ket \bra n|  
+ \sum^{\infty}_{n= a} \Bigl(\alpha |n\!+\!1\ket \bra n|+  
\Delta 
|n\ket \bra n|\Bigr) \,,  
\end{eqnarray}    
where we take $\Delta$ real using the rotational symmetry 
in the $(x,y)$ space. 
Here ${\cal Z}$ and ${\cal X}_+$ are  
$(a+b+1)\times (a+b+1)$ matrices with basis  
spanned by $|\!-\!b\ket,|\!-\!b\!+\!1\ket, \cdots,|a\!-\!1\ket, |a\ket$.  
We again  
insert this ansatz to the Gauss law and find 
\begin{eqnarray}    
[{\cal X}_+,[{\cal X}_-, 
{\cal Z}]] + [{\cal X}_-,[  {\cal X}_+,{\cal Z}]]   
=2l\alpha^2 |a\ket \bra a| 
-2l\rho^2 |\!-\!b\ket \bra -b\,|-\Delta Q 
\,,  
\label{masterdis}    
\end{eqnarray}    
where   
\begin{eqnarray}    
\Delta Q=\Delta({\cal Z}-l a){\cal X}_+ 
|a\ket \bra a| + \Delta|a\ket \bra a| {\cal X}_+ 
({\cal Z}-l a) 
+ ({\rm h.c.}) 
\,.  
\end{eqnarray}   
By the Gauss law, the boundary conditions  
for ${\cal Z}$ and ${\cal X}_+$ are set by  
\begin{eqnarray}    
\bra -b\, |{\cal Z}|\!-\!b\ket = -lb\,,\ \ \  
\bra a| {\cal Z}  |a\ket= l a\,,  
\label{conditiond1}    
\end{eqnarray}    
and  
\begin{eqnarray}    
\Bigl({\cal Z}+l(b\!-\!1)\Bigr){\cal X}_+|\!-\!b\ket =0\,,\ \ \  
 \Bigl({\cal Z}-l(a\!-\!1)\Bigr){\cal X}_-|a\ket=0\,.   
\label{conditiond2}    
\end{eqnarray}   
The left side of Eq.~(\ref{masterdis}) and $\Delta Q$  
are traceless. This implies that  
\begin{eqnarray}    
l \rho^2=l \alpha^2\,,   
\end{eqnarray}   
which  states that the number of  
fundamental strings is preserved through the bend. 
Let us consider the case  $a=1$ and $b=0$. The resulting most general 
configuration is 
\begin{eqnarray}    
{\cal Z} = l|1\ket\bra 1|\,, \ \  
{\cal X}_+ = {\rho\over \sqrt{2}}(|0\ket\bra 1|- |1\ket\bra 0|) 
\end{eqnarray} 
where again we set some gauge trivial phase factors to zero. 
This bend with displacement is depicted in Figure~2.
There is essentially no free parameter in this case. 
The shifted energy $\tilde{E}$ is evaluated as 
\begin{eqnarray}    
\tilde{E}={1\over 2} \rho^2(\rho^2+ \Delta^2)\,. 
\end{eqnarray}     
Contrary to the naive expectation,  
the energy does not approach to zero when $\Delta=0$. This 
is because there is a discontinuity at $\Delta=0$. 
Namely, when $\Delta=0$, there  
appears a new parameter. 

\begin{figure}
\epsfxsize=6.0in
\centerline{
\epsffile{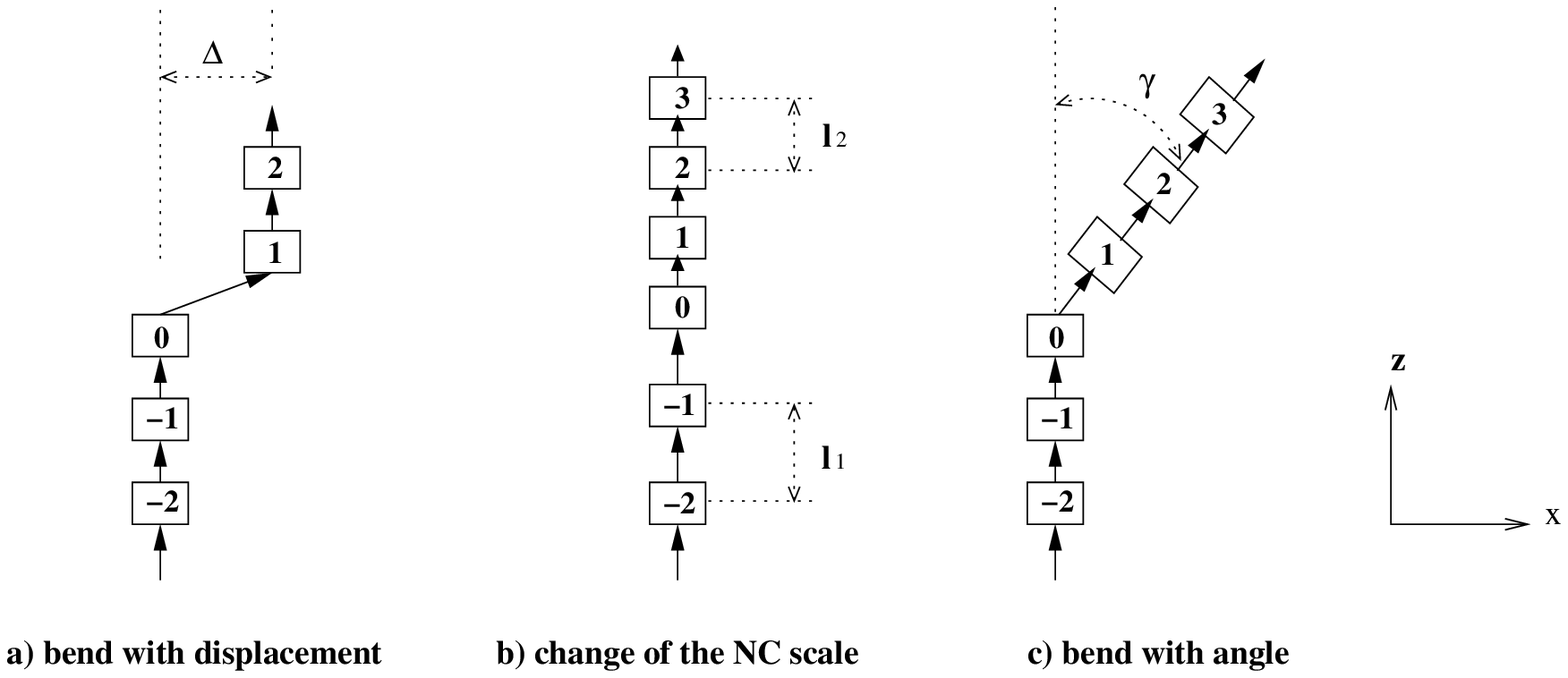}
}
\vspace{.1in}
{\small Figure~2:~A skeleton graph for the bend with displacement
appears first. The second represents a tube configuration along which the 
noncommutativity scales  changes at $z=0$.
The last graph is for the bend with angle.
}
\end{figure}
  
Now let us consider the following configurations. 
We make the bend of total displacement $\Delta$  
by repeating $k$ times of the unit bend with  
displacement $\Delta/k$.  
The unit bend with displacement $\Delta/k$ will cost  
the energy $\tilde{E}$ by $(1/2) \rho^2(\rho^2+ \Delta^2/k^2)$. 
The total cost of energy will then be 
\begin{eqnarray}    
\tilde{E}_{k}={1\over 2} \rho^2(k \rho^2+ \Delta^2/k)\,. 
\label{k=1}    
\end{eqnarray}   
If $k$ were an arbitrary  real number, 
the value 
\begin{eqnarray}    
\tilde{E}_{\rm min}=\rho^3 \Delta \,\,
\label{estimation}   
\end{eqnarray}   
would be the minimum at $k= \Delta/\rho$.  
Thus it appears that the bend with displacement
might be stabilized with its excited region kept finite.
We shall, however, prove below that this is not the case.

To verify (\ref{k=1}) for $k=2$, 
let us consider the case of 
$a=2$ and $b=0$ for a given nonvanishing real positive $\Delta$.
They are given by 
\begin{eqnarray}    
{\cal Z} \ &=& 2l|2\ket\bra 2| +l|1\ket\bra 1|\,, \nonumber\\  
{\cal X}_+ &=& \delta\, |2\ket\bra 2|+  
{\rho\over \sqrt{2}}\Bigl(e^{i\theta_1} |0\ket\bra 1| -  
e^{-i\theta_1}  |0\ket\bra 1|\Bigr) 
+{\rho\over 
\sqrt{2}}\Bigl(e^{i\theta_2} |0\ket\bra 2|- e^{-i\theta_2}   |2\ket\bra 0|\Bigr)\,, 
\end{eqnarray} 
where $\theta_1$ and $\theta_2$ are respectively the arguments of 
$\delta$ and $\Delta-\delta$. Here 
again we set some gauge trivial phase factors to zero. 
The shifted energy $\tilde{E}$ is given by 
\begin{eqnarray}    
\tilde{E}={1\over 2} \rho^2\Bigl(2\rho^2+ |\delta|^2+|\Delta-\delta|^2 +
\rho^2\sin^2(\theta_1-\theta_2)\Bigr)\,. 
\end{eqnarray}     
The minimum is achieved when $\theta_1=\theta_2=0$ and $\delta=\Delta/2$
with its value
\begin{eqnarray}    
\tilde{E}={1\over 2} \rho^2\Bigl(2\rho^2+ {|\Delta|^2\over 2}\Bigr)\,, 
\end{eqnarray}      
which agrees with (\ref{k=1}) for $k=2$. 

Another basic operation is the change of the noncommutativity scale 
within a tube. The configuration starts with the noncommutativity scale
$l_1$ for $z<0$ and becomes a tube with
 the noncommutativity scale
$l_2$ for $z>0$. One simple example is described by 
\begin{eqnarray}    
&& Z\,\,= 
l_1\sum^{-1}_{n=-\infty} n|n\ket \bra n| 
+l_2\sum^{\infty}_{n=0} n|n\ket \bra n| 
\,,  
\nonumber\\   
&&  X_+=  \rho_1\sum^{-1}_{n=-\infty} |n+1\ket \bra n| 
+\rho_2\sum^{\infty}_{n=0} |n+1\ket \bra n|\,, 
\end{eqnarray}    
which we illustrate in Figure~2 and Figure~3.
One may easily verify that this satisfies the Gauss law
constraint when $l_1\rho^2_1=l_2\rho^2_2$.
This  implies that the number of fundamental 
strings is preserved through the change of 
noncommutative scale. 
The excitation energy for this configuration reads
\begin{eqnarray}    
\tilde{E}={1\over 8} (\rho_1^2-\rho_2^2)^2\,.
\end{eqnarray}

\begin{figure}
\epsfxsize=4.1in
\centerline{
\epsffile{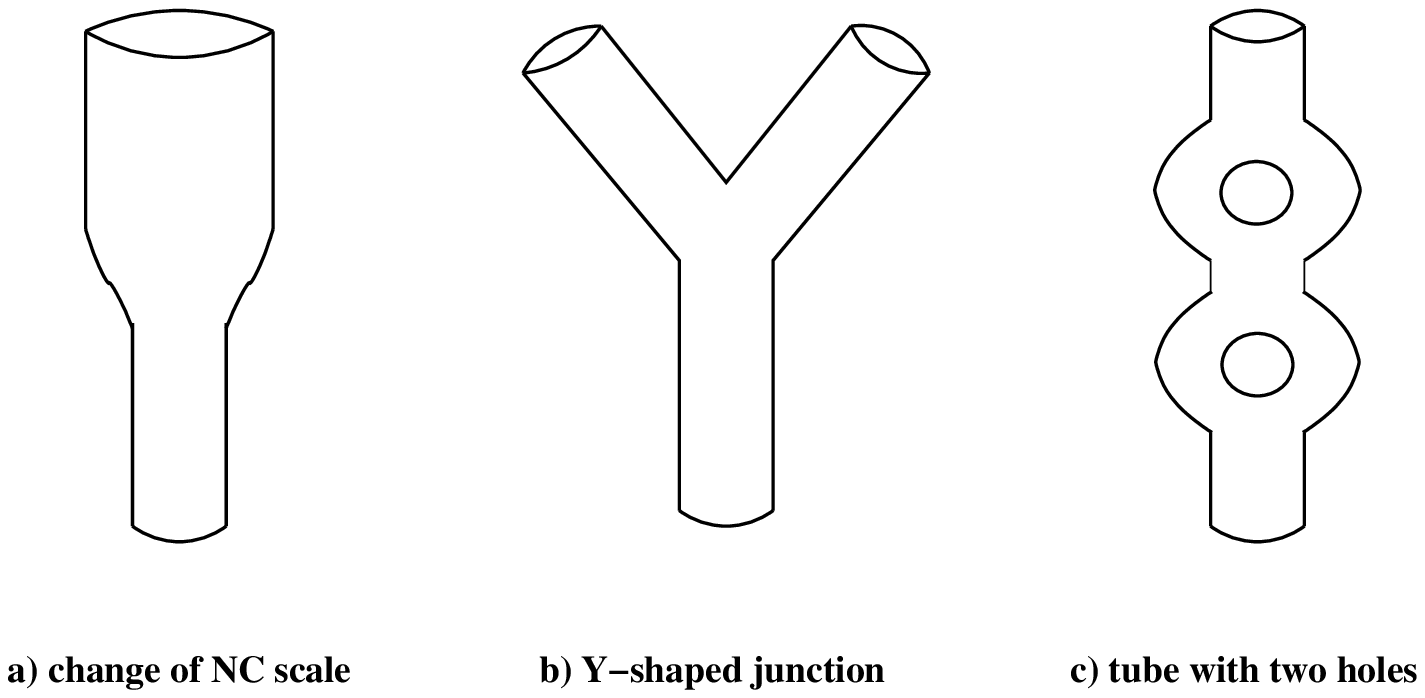}
}
\vspace{.1in}
{\small Figure~3:~The shape of tube, along which the noncommutativity
scale changes, is illustrated in the first figure. 
The second represents a Y-shaped junction.
The last picture is for a tube  with two holes.
}
\end{figure}

We now discuss the bend with angle. First let 
us note that there are rotational symmetries
in the target space $(X,Y,Z)$. For example, a 
rotation by an angle $\gamma$ in the $(X,Z)$ plane 
may produce another
supersymmetric tube if one starts from a
tube with its axis along 
the z-direction. Choosing a gauge 
$A_0=\cos\gamma Z +\sin\gamma X$, the rotated supersymmetric
 tube solution is given by
\begin{eqnarray}    
&& Z\,\,=\cos\gamma  
\sum^{\infty}_{n=-\infty}l n|n\ket \bra n| 
-\sin\gamma  \sum^{\infty}_{n=-\infty}
{\rho\over 2}(|n+1\ket \bra n|+|n\ket \bra n+1|)
\,,  
\nonumber\\   
&&  X=  
\cos\gamma \sum^{\infty}_{n=-\infty}{\rho\over 2} 
(|n+1\ket \bra n|+|n\ket \bra n+1|)+
\sin\gamma  
\sum^{\infty}_{n=-\infty}l n|n\ket \bra n| 
\,,\nonumber\\  
&&  Y=  \sum^{\infty}_{n=-\infty}
{\rho\over 2i}(|n+1\ket \bra n|-|n\ket \bra n+1|)\,.
\end{eqnarray}    
To obtain the bended configuration with angle, 
we shall combine the above solution with the supersymmetric 
tube with $\gamma=0$. For this purpose, we shall work 
in a gauge $A_0=\sum^{\infty}_{n=-\infty}l n|n\ket \bra n|$.
The Gauss law for a static configuration will then be
\begin{eqnarray}    
[X_i,[X_i, A_0]]=0\,.
\end{eqnarray}    
It is straightforward to verify that the following
configuration
\begin{eqnarray}    
&& Z = 
\sum^{-1}_{n=-\infty}l n|n\ket \bra n| 
+\cos\gamma  
\sum^{\infty}_{n=0}l n|n\ket \bra n| 
-\sin\gamma  \sum^{\infty}_{n=0}
{\rho\over 2}(|n+1\ket \bra n|+|n\ket \bra n+1|)
\,,  
\nonumber\\   
&&  X=  
\sum^{-1}_{n=-\infty} {\rho\over 2}(|n\!+\!1\ket \bra n|\!+
\!|n\ket \bra n\!+\!1|)
+
\cos\gamma  \sum^{\infty}_{n=0}
{\rho\over 2}(|n\!+\!1\ket \bra n|\!+\!|n\ket \bra n\!+\!1|)
+\sin\gamma  
\sum^{\infty}_{n=0}l n|n\ket \bra n| 
\,,\nonumber\\  
&&  Y=  \sum^{\infty}_{n=-\infty}
{\rho\over 2i}(|n+1\ket \bra n|-|n\ket \bra n+1|)
\end{eqnarray}    
does satisfy the Gauss law. In this 
configuration, the tube is along the z-direction  for $z<0$
while its axis is in the 
$(\sin\gamma,0,\cos\gamma)$-direction for $z>0$. 
The skeleton graph for this configuration is depicted in Figure~2.
The bend occurs 
at $z=0$. 
Its excitation energy 
may be computed 
using 
\begin{eqnarray}    
\tilde{E}= {1\over 2} \tr \left(\partial_0 X^i \partial_0 X^i 
-[X,Y]^2-[X,Z]^2-[Y,Z]^2+[A_0,X^i][A_0,X^i]\right)     
\,. 
\end{eqnarray}
This energy density is also conserved in time and matches with
${\cal H}- C_J$ asymptotically.  
Though complicated, the straightforward evaluation
of the energy for the bend
 gives
 \begin{eqnarray}    
\tilde{E}={\rho^4\over 16}\Bigl(16-(3+\cos\gamma)^2\Bigr) 
\,. 
\end{eqnarray}
For $\gamma=0$, the tube is straight and 
the corresponding
excitation energy vanishes as it should be.
The excitation energy increases as the angle grows
and reaches its maximum for $\gamma=\pi$.
For finite $\gamma$, we expect that the excitation 
 energy may be lowered by making a bend gradually.
However, we shall not explore such  
configurations.
Finally, we like to comment on the bend with a 
displacement formed by combining double bends 
with angles. Namely it is achieved by
one bending at $z=0$ with  angle $\gamma$ and the 
other bending at $z'=lk$ with angle 
$-\gamma$. The corresponding displacement will be
$\Delta= l k \sin\gamma$. For a given $\Delta$, the excitation
energy may be brought to arbitrarily small 
by sending $k$ to large with $\gamma\sim \Delta/(lk)$.
Hence we see that the bend with displacement
cannot be stabilized with a finite range of the excited region.

\begin{figure}
\epsfxsize=2.7in
\centerline{
\epsffile{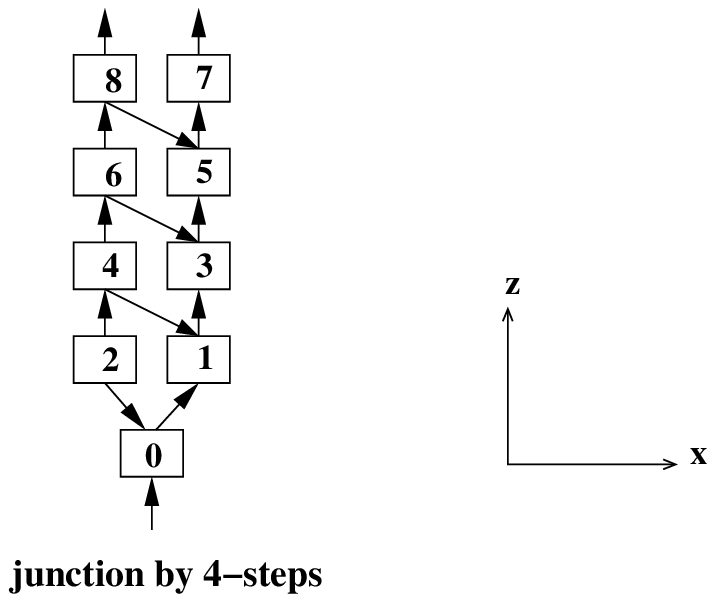}
}
\vspace{.1in}
{\small Figure~4:~A junction constructed by $k$ steps  
repeatedly using  splitting and joining. 
In the splitting or joining,
the smaller $ {(radius)}^2$ is given by $N_\beta/(kl)$.
Here we illustrate the case of $k=4$ for simplicity.
}
\end{figure}

At this point let us discuss energetics involved with
the junction constructed previously. As in Figure~3,
one may construct the junction by $k$ steps. In this case,
the corresponding total excitation energy will be
 \begin{eqnarray}    
\tilde{E}_k={g_s l_s\over4 l^2} N_\beta \left(N_\rho +{N_\beta
\over k^2}(11k-6)
\right) 
\,. 
\end{eqnarray}
Since $\tilde{E}_k$ monotonically decreases as the integer $k$ grows,
it appears that the minimum occurs at $k=\infty$. However, $k$
cannot exceed $N_\beta$ due to the quantization of 
$l\, {(radius)}^2$. Hence the estimation of the minimum is
 \begin{eqnarray}    
\tilde{E}_{\rm min}={g_s l_s\over4 l^2} 
N_\beta \left(N_\rho +11-{6\over N_\beta}
\right) 
\,. 
\end{eqnarray} 
The corresponding configuration is localized over a range 
$\delta z\sim l N_\beta$.
Such  stabilization of the localized configuration  
occurs 
due 
to the semiclassical quantization of radius.

As we discussed in detail, the basic construct for the
general Riemann surfaces will be the junction, the bend
with angle and the connector with change of the noncommutativity scale. 
 These basic local operations may 
be combined
to produce arbitrarily 
complicated configurations with varying noncommutativity scale.
These general configurations resemble  string loop diagrams 
but everything occurs in spatial dimensions.

The number of fundamental strings stretched and the number 
of D0's play an important role in the understanding
of such general configurations.
The sum of the cross sectional area multiplied by the 
noncommutativity scale is preserved in any case. As we have 
shown above, this quantity is proportional to the 
total number of
the stretched fundamental string 
along the tube. The D0 brane density along the tube is given by
the inverse of the noncommutativity scale $l$. 
For the given number of
the stretched fundamental strings $N_s$, 
the radius of the tube gets
larger if the D0's are densely packed.  
Namely $N_s$ and the density of D0-branes control 
the radius of tubes.

\section{Conclusions}
We have constructed the junction that separates
one tube to two. One may also use this local operation to join 
two tubes to one. If one uses the junctions repeatedly,
one may split one tube to arbitrary number of tubes or join 
arbitrary number of tubes to one. The operation of bend with angle
makes a tube direct to an arbitrary spatial direction. 
If one combines  the junction and the bend operation, one may 
construct
a Y-shaped junction in Figure~3.  
The basic constructs are only locally 
excited, so are the combinations.
For example,
the configuration with 
two holes depicted in Figure~3 
may be constructed by 
splitting a tube to two, adding appropriate bending, 
joining the two to one and repeating the 
whole operations again. 
In conclusion, one may have arbitrary Riemann surfaces by combining
two basic operations. 
Furthermore if one adds  noncommutativity-changing operations,
one may have a varying noncommutativity over the Riemann surface.

From the view point of worldvolume gauge theories,
the physical meaning of junctions is quite intriguing.
When 
$\alpha=\beta=\rho/\sqrt{2}$, the junction in (\ref{ansatz})  
interpolates the U(1)  noncommutative theory
and the U(2) noncommutative gauge theory. Hence the junction works 
as a kind of domain wall. Similarly by considering a tube splitting 
into many tubes of an equal radius, one finds that a domain wall 
interpolating the  U(1) noncommutative theory to U(p) 
noncommutative gauge theory. Of course, the original matrix theory 
governs all these dynamics.

In this note, we first review the previous construction 
of supersymmetric tubes 
in the matrix model.
We identified all the charges and the moment carried by
the supersymmetric tubes.
The symmetry breaking of many tubes with different radii is discussed. 
We then study the fluctuation spectra for 
the tube-D0 system. 
 We   
construct more general physical configurations of tubes
that are consistent with the Gauss law. The basic 
constructs are shown to be the junction, the 
bend with angle and the connector of tubes with different noncommutativity 
scales. 
Combining these basic operations 
one may construct even arbitrary  Riemann surfaces with arbitrary
topology. The noncommutativity scale may vary over the Riemann surfaces.

The tubes involve the background electric component that is 
critical. For the worldvolume theory, we have utilized only the 
spatial noncommutativity. The possible role of the electric field 
for the spacetime noncommutativity is not fully understood.
In particular, the only invariant combination in the 2+1 dimensions
is 
$(E^2-B^2/\rho^2)(2\pi\alpha')^2= 1-(2\pi\alpha')^2/(l\rho)^2$.
If one sends $\alpha'$ to zero while fixing  $l\rho$, this corresponds
to the NCOS limit\footnote{This observation is due to 
S. Minwalla}\cite{Gopakumar2}. 
Detailed investigation on the nature
of the limit is necessary.  


Finally the dynamical issues of the tubular configurations 
 are not clearly understood. There is no 
intuitive understanding why there is no force between 
many parallel tubes. There is no force between the 
tube and D0
and again the intuitive understanding is lacking.
One of approach in this direction would be the study of the
worldsheet CFT of $p\!-\!p'$ strings as in Ref.\cite{Chen}.
Detailed dynamical investigation of all kind of the configurations will
be quite interesting. However, due to complications, one  needs
better methods of organizing such dynamical processes.
%
Further studies are required in this direction.

\noindent{\large\bf Acknowledgment} 
We are especially benefitted from  the extensive 
discussions with Kimyeong Lee at the early stage of 
this work. 
DB would like to thank J.-H. Cho, R. Gopakumar,  M. Gutperle,
A. Karch, Taejin Lee, S. Minwalla
 and A. Strominger for the enlightening discussions.
DB also likes to 
acknowledge the Harvard Theory Group 
 where part of this work was done. 
This work is supported in part by 
KOSEF 1998 Interdisciplinary Research Grant 98-07-02-07-01-5  
and by UOS Academic Research Grant. 
  


\end{document}